\documentclass[showpacs,amsmath,amssymb]{revtex4}

\usepackage{graphicx}
\usepackage{dcolumn}
\usepackage{bm}
\usepackage{epsfig}

\begin{document}

\def\0#1#2{\frac{#1}{#2}}
\def\bct{\begin{center}} \def\ect{\end{center}}
\def\beq{\begin{equation}} \def\eeq{\end{equation}}
\def\bea{\begin{eqnarray}} \def\eea{\end{eqnarray}}
\def\nnu{\nonumber}
\def\n{\noindent} \def\pl{\partial}
\def\g{\gamma}  \def\O{\Omega} \def\e{\varepsilon} \def\o{\omega}
\def\s{\sigma}  \def\b{\beta} \def\p{\psi} \def\r{\rho}
\def\G{\Gamma} \def\S{\Sigma} \def\l{\lambda}

\title{the CJT calculation in studying nuclear matter beyond mean field approximation}
\author{Song~Shu}
\affiliation{Faculty of Physics and Electronic Technology, Hubei
University, Wuhan 430062, China}
\author{Jia-Rong~Li}
\affiliation{Institute of Particle Physics, Hua-Zhong Normal
University, Wuhan 430079, China}
\begin{abstract}
We have introduced a CJT calculation in studying nuclear matter
beyond mean field approximation. Based on the CJT formalism and
using Walecka model, we have derived a set of coupled Dyson
equations of nucleons and mesons. Neglecting the medium effects of
the mesons, the usual MFT results could be reproduced. The beyond
MFT calculations have been performed by thermodynamic consistently
determining the meson effective masses and solving the coupled gap
equations for nucleons and mesons. The numerical results for the
nucleon and meson effective masses at finite temperature and
chemical potential in nuclear matter are discussed.
\end{abstract}  \maketitle

\section{Introduction}
Mean field theory (MFT) has an important role in the study of
nuclear matter. Based on quantum hadrodynamics (QHD) and by mean
field approximation, one can obtain the results in good agreement
with the properties of the nuclear matter, like saturation density
and binding energy, etc~\cite{ref1,ref2,ref3}. For Walecka model
(QHD-I), mean field approximation could be regarded as replacing
the meson fields by their expectation value~\cite{ref4}. That is
to say treating the meson fields as the classical fields while
quantum fluctuation ignored. From a Green's function formalism,
the mean field approximation could also be viewed as resumming all
the tad-pole graphs in the nucleon self-energy~\cite{ref5,ref6}.
If one wants to make calculations beyond MFT, one should
systematically calculate the higher loop contributions from a
self-consistent resummation scheme. At the same time it is
important to preserve the thermodynamic consistency~\cite{ref6a}.
By the one particle irreducible effective action formalism, when
evaluated to two loops, the calculations become very involved and
after taking certain approximations the results are found to be
not consistent with the properties of the nuclear
matter~\cite{ref7}. In principle the higher loop contributions
should give reasonable results. Therefore one should find
systematic and consistent resummation scheme and make proper
approximations to obtain soundable results beyond mean field
calculation.

For the effective action formalism, the effective action $\Gamma
(\phi)$, which is the generating functional of one particle
irreducible graphs, could been systematically calculated by
summing all the relevant Feynman graphs to a given order of the
loop expansion~\cite{ref8,ref9,ref10,ref11}. However in this
scheme it is not convenient to preserve the consistency of the
effective action and the full propagators at the same order of the
approximation. A generalized effective action for composite
operators has been developed by Cornwall, Jackiw and Tomboulis
(CJT)~\cite{ref12}. According to CJT formalism the usually
effective action $\G(\phi)$ is generalized to depend not only on
$\phi(x)$, a possible expectation value of the quantum field
$\Phi(x)$, but also on $G(x,y)$, a possible expectation value of
the time-ordered product $T\Phi(x)\Phi(y)$. $G(x,y)$ is also the
full propagator of the field. In this case the effective action
$\G(\phi,G)$ is the generating functional of the two particle
irreducible graphs. An effective potential $V(\phi,G)$ can be
defined by removing an overall factor of space-time volume of the
effective action. Physical solutions demand minimization of the
effective potential with respect to both $\phi$ and
$G$~\cite{ref12}, which means \bea \0{\pl V(\phi,G)}{\pl \phi}=0
\eea and \bea\label{eq1} \0{\pl V(\phi,G)}{\pl G}=0. \eea This
formalism was originally written at zero temperature. Then it had
been extended at finite temperature by Amelino-Camelia and Pi
where it was used for investigations of the effective potential of
the $\lambda\phi^4$ theory~\cite{ref13} and gauge
theories~\cite{ref14}. In the study of the spontaneous symmetry
breaking system, $\phi$ is also the order parameter of the
transition and nonzero; while in the system without spontaneous
symmetry breaking, this parameter is zero. For the thermodynamic
system the effective potential could be identified as the
thermodynamic potential. One can use the CJT formalism to study
nuclear matter. In our previous work~\cite{ref15}, we have already
applied the CJT formalism to the study of nuclear matter. At the
first step, after the proper approximations we have successfully
reproduced the MFT results through the CJT formalism. And we have
also illustrated that the thermodynamic consistency is preserved
in the CJT formalism. In this paper we will further study how to
make thermodynamic consistent calculations beyond MFT in the CJT
formalism.

The organization of the present paper is as follows: In section 2,
we formulate the CJT formalism in the study of nuclear matter by
the Walecka model. The thermodynamic potential and coupled Dyson
equations of the full propagators are derived at Hartree-Fock
approximation. In section 3, we will make the beyond MFT
calculations by including the medium effects of the mesons in a
thermodynamic consistent way and give our numerical results. The
last section comprises a summary and discussion.

\section{CJT formalism in nuclear matter}
We start from the QHD-I Lagrangian which can be written as \bea
{\cal L}=\bar\p(i\g_\mu\pl^\mu-m_N+g_\s\s-g_\o\g_\mu\o^\mu)\p
+\012(\pl_\mu\s\pl^\mu\s-m_\s^2\s^2) -\014F_{\mu\nu}F^{\mu\nu}+
\012m_\o^2\o_\mu\o^\mu, \eea where $\p$, $\s$ and $\o$ are the
fields of the nucleon, sigma meson and omega meson respectively
and $F_{\mu\nu}=\pl_\mu\o_\nu-\pl_\nu\o_\mu$. According to the
Lagrangian we can write down the free propagators of the nucleon,
sigma meson and omega meson respectively as \bea
G_0(p)=\0i{/\!\!\!p-m_N}, \\ \Delta_0(p)=\0i{p^2-m_\s^2}, \\
D_0^{\mu\nu}(p)= \0{-ig^{\mu\nu}}{p^2-m_\o^2}. \eea

We will use the imaginary-time formalism to compute quantities at
finite temperature~\cite{ref16}. Our notation is \bea \int_p
f(p)\equiv\0i{\b}\sum_n\int\0{d^3\bf p}{(2\pi)^3}f(p_0=i\o_n,{\bf
p}) , \eea where $\b$ is the inverse temperature, $\b=1/T$. We
have $\o_n=2n\pi T$ for boson and $\o_n=(2n+1)\pi T$ for fermion,
where $n=0, \pm1, \pm2, \cdots$. A baryon chemical potential $\mu$
can be introduced by replacing $p_0=i\o_n$ with $p_0=i\o_n+\mu$ in
the nucleon propagator.

\begin{figure}[tbh]
\begin{center}
\includegraphics[width=210pt,height=90pt]{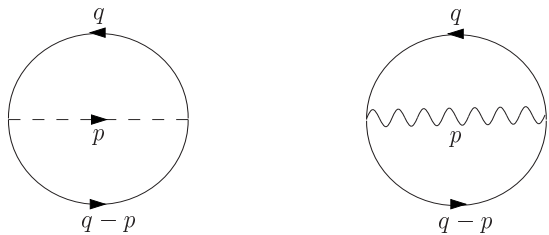}
\end{center}
\caption{Hartree-Fock approximation to $\O_2(G,\Delta,D)$. The
solid, dash and wavy lines are the nucleon propagator G, the $\s$
meson propagator $\Delta$ and the $\o$ meson propagator $D$
respectively. $q$ and $p$ are the four momenta.}\label{x1}
\end{figure}

\begin{figure}[tbh]
\begin{center}
\includegraphics[width=250pt,height=160pt]{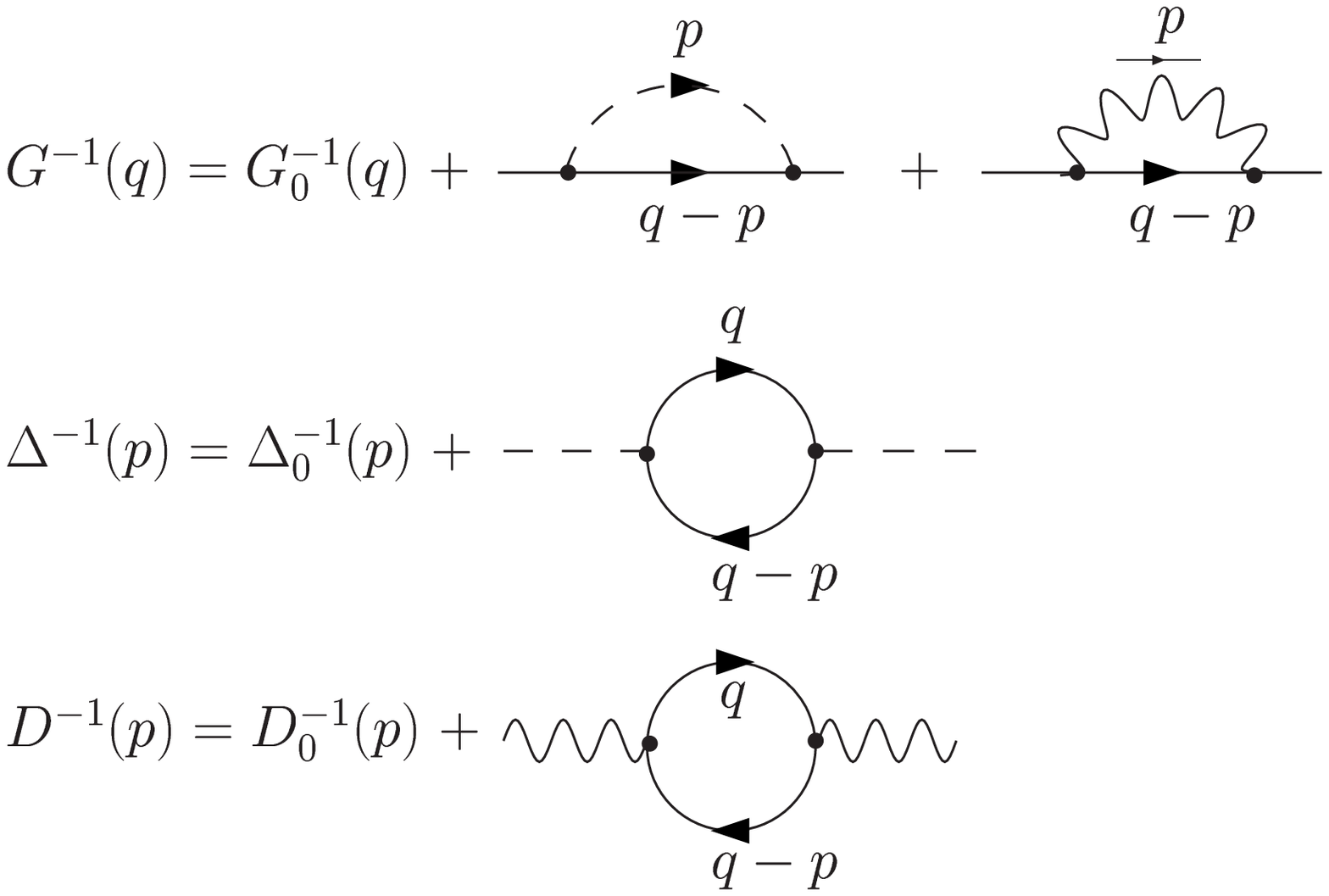}
\end{center}
\caption{The Dyson equations satisfied by the nucleon, the $\s$
meson and the $\o$ meson propagators at Hartree-Fock
approximation. $q$ and $p$ are the four momenta.}\label{x2}
\end{figure}

According to the CJT formalism~\cite{ref12}, the expansion of the
effective potential or the thermodynamic potential of the nuclear
matter can be written as \bea \O(G, \Delta, D)=i\int_pTr\ln
G_0(p)G^{-1}(p)&+&i\int_pTr\left
[G_0^{-1}(p)G(p)-1\right]-\0i2\int_p\ln\Delta_0(p)\Delta^{-1}(p)-
\0i2\int_p \left[\Delta_0^{-1}(p)\Delta(p)-1\right] \nnu
\\ &-&\0i2\int_p\ln
D_0(p)D^{-1}(p)-\0i2\int_p\left[D_0^{-1}(p)D(p)-1\right]+\O_2(G,\Delta,D),
\eea where $G$, $\Delta$ and $D$ are the full propagators of
nucleon, $\s$ meson and $\o$ meson respectively, which are
determined by the stationary condition (\ref{eq1}).
$\O_2(G,\Delta,D)$ is given by all the two-particle irreducible
vacuum graphs with all the propagators treated as the full
propagators. In the CJT formalism at Hartree-Fock approximation
$\O_2(G,\Delta,D)$ can be illustrated by the graphs of figure
\ref{x1}, where the solid lines repent $G(p)$, the dashed line
represents $\Delta(p)$ and the wavy line represents
$D^{\mu\nu}(p)$. The vertices for $\s\p\bar\p$ and
$\o^\mu\p\bar\p$ are $-g_\s$ and $g_\o\g^\mu$ respectively.
The analytic expression is \bea
\O_2(G,\Delta,D)=\0{ig_\s^2}2\int_p\int_q
Tr\left[G(q)G(q-p)\Delta(p)\right]+\0{ig_\o^2}2\int_p\int_q
Tr\left[\g^\mu G(q)\g^\nu G(q-p)D_{\mu\nu}(p)\right]. \eea From
the stationary condition (\ref{eq1}) which demands that
$\O(G,\Delta,D)$ be stationary against variations of $G$, $\Delta$
and $D$ respectively, we will have the following Dyson equations
\bea G(q)^{-1}&=&G_0(q)^{-1}+g_\s^{2}\int_{p}G(q-p)\Delta(p)+
g_\o^2\int_{p}\g^\mu G(q-p)\g^\nu D_{\mu\nu}(p), \label{eq2} \\
\Delta(p)^{-1}&=&\Delta_0(p)^{-1}-g_\s^2\int_q
Tr\left[G(q)G(q-p)\right], \label{eq2a} \\
D_{\mu\nu}(p)^{-1}&=&D_{0, \mu\nu}(p)^{-1}-g_\o^2\int_q
Tr\left[\g_\mu G(q)\g_\nu G(q-p)\right]. \label{eq2b} \eea The
above equations can be also represented pictorially in figure
\ref{x2}.
These integration equations are nonlinear coupled and momentum
dependent. Needless to say they are very difficult for
computation. The certain approximations should be adopted.

\section{the beyond mean field CJT calculation and numerical results}
\begin{figure}[tbh]
\begin{center}
\includegraphics[width=220pt,height=140pt]{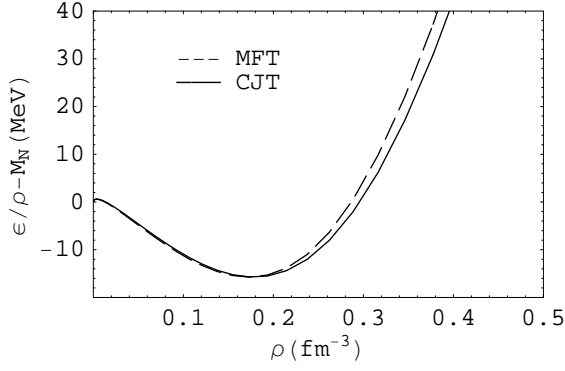}
\end{center}
\caption{The average energy per nucleon minus the nucleon mass
$m_N$ as functions of baryon density at zero temperature in the
CJT formalism (solid line) and MFT (dashed line).}\label{x3}
\end{figure}

In our previous study~\cite{ref15}, in order to reproduce the mean
field results through the CJT formalism, we have neglected the
medium effects of the mesons which means the meson full
propagators are replaced by their bare ones. The Dyson equations
are decoupled and can be numerically solved. Then we could
reproduce the correct saturation properties of nuclear matter as
shown in figure \ref{x3}. The coupling constants thus are fixed at
$g_\s^2=155.6$ and $g_\o^2=521.5$. The result from MFT is also
presented for comparison in figure \ref{x3}.

In this paper, we are going to make calculations beyond mean
field. We will take the following ansatz of the meson full
propagators \bea \Delta(p)=\0i{p^2-M_\s^2}, \\
D^{\mu\nu}(p)= \0{-ig^{\mu\nu}}{p^2-M_\o^2}, \eea where $M_\s$ and
$M_\o$ are the effective masses of the $\s$ and $\o$ mesons, which
in principle should be determined by equation (\ref{eq2a}) and
(\ref{eq2b}). However, as we have mentioned that the momentum
dependent coupled Dyson equations are very difficult to be solved.
So we have to find other simplified ways to determine the
effective masses and meanwhile preserve the thermodynamic
consistency.

The full nucleon propagator is taken as the following form
\bea\label{eq3} G(q)=\0i{\g_{\mu}q^{\mu}-m_N+\S}, \eea where $\S$
is the proper nucleon self-energy which will be determined by the
equation (\ref{eq2}). It can be generally written as
\bea\label{eq4} \S(q)=\S_s(q)-\g^0\S_0(q)+{\bm\g}{\bm\cdot}{\bf
q}\S_v(q). \eea Thus we can define an effective nucleon mass
\bea\label{eq4a} M_N(q)=m_N-\S_s(q), \eea which is momentum
dependent. As the nuclear matter is a uniform system at rest, the
contribution of $\S_v$ term in equation (\ref{eq4}) is very small
and usually neglected~\cite{ref5}. So only $\S_s$ and $\S_0$ need
to be evaluated.

By the above ansatz of the full propagators, the thermodynamic
potential could be written as \bea \O(G, \Delta, D)=&i&\int_pTr\ln
G_0(p)G^{-1}(p)-\int_pTr\left
[G(p)\S(p)\right]-\0i2\int_p\ln\Delta_0(p)\Delta^{-1}(p)-
\012\int_p \left(M_\s^2-m_\s^2\right)\Delta(p) \nnu
\\ &-&\0i2\int_p\ln
D_0(p)D^{-1}(p)+\012\int_p\left(M_\o^2-m_\o^2\right)g_{\mu\nu}D^{\mu\nu}(p)+\O_2(G,\Delta,D).
\label{eq5} \eea From equation (\ref{eq2}), (\ref{eq3}) and
(\ref{eq4}), we could derive that \bea
\S(q)=ig_\s^2\int_pG(q-p)\Delta(p)+ig_\o^2\int_p\g_\mu
G(q-p)\g_\nu D^{\mu\nu}(p), \label{eq6} \eea thus we have \bea
\O_2(G,\Delta,D)&=&\012Tr\int_qG(q)\left\{ig_\s^2\int_pG(q-p)\Delta(p)+ig_\o^2\int_p\g_\mu
G(q-p)\g_\nu D^{\mu\nu}(p)\right\} \nnu \\ &=& \012\int_qTr\left
[G(q)\S(q)\right]. \eea

In the next we will mainly calculate $\S$, $M_\s$, $M_\o$ and
$\Omega$ in a thermodynamic consistency way. In the following
calculations the vacuum fluctuations will be neglected. The
$\S(q)$ could be evaluated from equation (\ref{eq6}). The momentum
dependence can be eliminated if we define the nucleon effective
mass as the pole mass of the full propagator~\cite{ref16a} in the
limit ${\bf q}\to 0$ as discussed in~\cite{ref15}. With the $\S_v$
term dropped we have \bea \S&=&\S_s-\g^0\S_0, \\
\S_s&=&-g_\s^2\int \0{d^3\bf p}{(2\pi)^3} \0{M_N}{2E_\s
E}\0{2(M_N-E)n_\s-E_\s(\tilde n_++\tilde
n_-)}{(M_N-E)^2-E_\s^2}+g_\o^2\int \0{d^3\bf p}{(2\pi)^3}
\0{2M_N}{E_\o E}\0{2(M_N-E)n_\o-E_\o(\tilde
n_++\tilde n_-)}{(M_N-E)^2-E_\o^2}, \label{7a} \\
\S_0&=&g_\s^2\int \0{d^3\bf p}{(2\pi)^3} \0{-(\tilde n_+-\tilde
n_-)} {2(M_N-E)^2-2E_\s^2}+g_\o^2\int \0{d^3\bf p}{(2\pi)^3}
\0{-(\tilde n_+-\tilde n_-)}{(M_N-E)^2-E_\o^2}, \label{7b} \eea
where $E_{\s/\o}=\sqrt{{\bf p^2}+m_{\s/\o}^2}$ and $E=\sqrt{{\bf
p}^2+M_N^2}$; the Bose and Fermion distribution functions are \bea
n_{\s/\o}=\01{e^{\b E_{\s/\o}}-1}, \ \ \ \ \ \ \tilde
n_\pm=\01{e^{\b(E\mp\mu^*)}+1}, \eea where $\mu^*=\mu-\S_0$ and
$\mu$ is the baryon chemical potential.

After performing the frequency sums in equation (\ref{eq5}) and
neglecting the vacuum fluctuations, the thermodynamic potential
could be evaluated as \bea\label{eq8} \O(\mu,T)=&-&4T\int
\0{d^3\bf p}{(2\pi)^3}\left[\ln (1+e^{-\b(E-\mu^*)}) + \ln
(1+e^{-\b(E+\mu^*)})\right] \nnu \\ &+&4M_N\S_s\int \0{d^3\bf
p}{(2\pi)^3}\01E(\tilde n_++\tilde n_-)-4\S_0 \int \0{d^3\bf
p}{(2\pi)^3}(\tilde n_+-\tilde n_-) \nnu \\ &+&T\int \0{d^3\bf
p}{(2\pi)^3}\ln (1-e^{-\b E_\s})-\012(M_\s^2-m_\s^2)\int\0{d^3\bf
p}{(2\pi)^3}\0{n_\s}{E_\s} \nnu \\ &+&4T\int \0{d^3\bf
p}{(2\pi)^3}\ln (1-e^{-\b E_\o})-2(M_\o^2-m_\o^2)\int\0{d^3\bf
p}{(2\pi)^3}\0{n_\o}{E_\o}+\O_2, \eea where \bea
\O_2=-2M_N\S_s\int \0{d^3\bf p}{(2\pi)^3}\01E(\tilde n_++\tilde
n_-)+2\S_0 \int \0{d^3\bf p}{(2\pi)^3}(\tilde n_+-\tilde n_-).
\eea The $\S_s$ and $\S_0$ in $\O_2$ will be evaluated in the
forms of equations (\ref{7a}) and (\ref{7b}). Now we are in a
position to determine the effective masses $M_\s$ and $M_\o$. In
equation (\ref{eq8}), $M_\s$ and $M_\o$ could be viewed as
independent variables. As known that a thermodynamic system in
equilibrium will minimize its thermodynamic potential to these
independent variables, which means \bea \0{\pl\O}{\pl M_\s}=0, \ \
\ \ \ \0{\pl\O}{\pl M_\o}=0. \label{eq9} \eea $M_\s$ and $M_\o$
determined in this way are momentum independent and the
thermodynamic consistency is certainly preserved. From equation
(\ref{eq8}), the two equations in (\ref{eq9}) could be written as
\bea \012(M_\s^2-m_\s^2)\0{\pl}{\pl M_\s}\int\0{d^3{\bm
p}}{(2\pi)^3}\0{n_\s}{E_\s}+2M_N\int\0{d^3{\bm
p}}{(2\pi)^3}\01E(\tilde n_++\tilde n_-)\0{\pl\S_s}{\pl M_\s}
-2\int\0{d^3{\bm p}}{(2\pi)^3}(\tilde n_+-\tilde
n_-)\0{\pl\S_0}{\pl M_\s}=0, \label{10} \eea \bea
(M_\o^2-m_\o^2)\0{\pl}{\pl M_\o}\int\0{d^3{\bm
p}}{(2\pi)^3}\0{n_\o}{E_\o}+M_N\int\0{d^3{\bm
p}}{(2\pi)^3}\01E(\tilde n_++\tilde n_-)\0{\pl\S_s}{\pl M_\o}
-\int\0{d^3{\bm p}}{(2\pi)^3}(\tilde n_+-\tilde
n_-)\0{\pl\S_0}{\pl M_\o}=0. \label{11} \eea

The equations (\ref{7a}), (\ref{7b}), (\ref{10}) and (\ref{11})
form a set of closed equations from which $\S_s$, $\S_0$, $M_\s$
and $M_\o$ could be numerically solved. At this stage we could
make calculations and discussions beyond MFT in studying nuclear
matter. As an application we will discuss the temperature and
chemical potential dependence of the effective masses of nucleons,
$\s$ mesons and $\o$ mesons in the hot and dense nuclear matter.
\begin{figure}[tbh]
\begin{center}
\includegraphics[width=220pt,height=140pt]{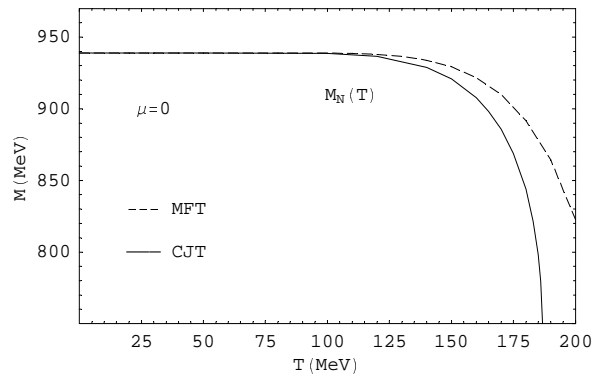}
\end{center}
\caption{The effective nucleon masses as functions of temperature
at zero chemical potential in the CJT formalism (solid line) and
MFT (dashed line).}\label{x4}
\end{figure}

\begin{figure}[tbh]
\begin{center}
\includegraphics[width=220pt,height=140pt]{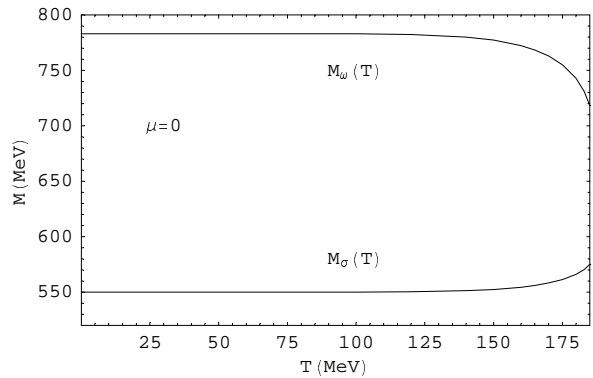}
\end{center}
\caption{The effective masses of $\s$ and $\o$ mesons as functions
of temperature at zero chemical potential in the CJT
formalism.}\label{x5}
\end{figure}

At zero chemical potential, we could discuss the temperature
dependence of the effective masses. In figure \ref{x4} the nucleon
effective mass of the CJT case decreases with temperature
increasing. At low temperature the effective mass decreases very
slowly. At certain high temperature the effective mass becomes
decreasing fast with temperature increasing. The MFT result is
also plotted for comparison. We could see that when temperature is
high the effective mass decreases more rapidly in the CJT case
than that of the MFT case. In figure \ref{x5} the meson effective
masses as functions of temperature are plotted. The effective mass
of $\s$ meson increases with temperature increasing. At high
temperature it increases more obviously. While The effective mass
of $\o$ meson will decrease with temperature increasing. It
decreases more quickly at high temperature.

\begin{figure}[tbh]
\begin{center}
\includegraphics[width=220pt,height=140pt]{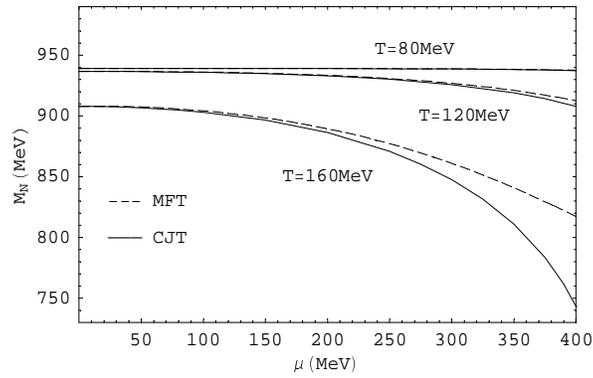}
\end{center}
\caption{The effective nucleon masses as functions of chemical
potential at different given temperatures in the CJT formalism
(solid line) and MFT (dashed line).}\label{x6}
\end{figure}

At given temperature, we could also study the chemical potential
dependence of the effective masses. In figure \ref{x6} we could
see that the nucleon effective masses decrease with chemical
potential increasing at different given temperatures. From low
temperature to high temperature the effective masses decrease more
and more remarkably with chemical potential increasing. In the CJT
case the effective masses decrease more rapidly with chemical
potential than that in the MFT case at high temperatures.

\begin{figure}[tbh]
\begin{center}
\includegraphics[width=220pt,height=140pt]{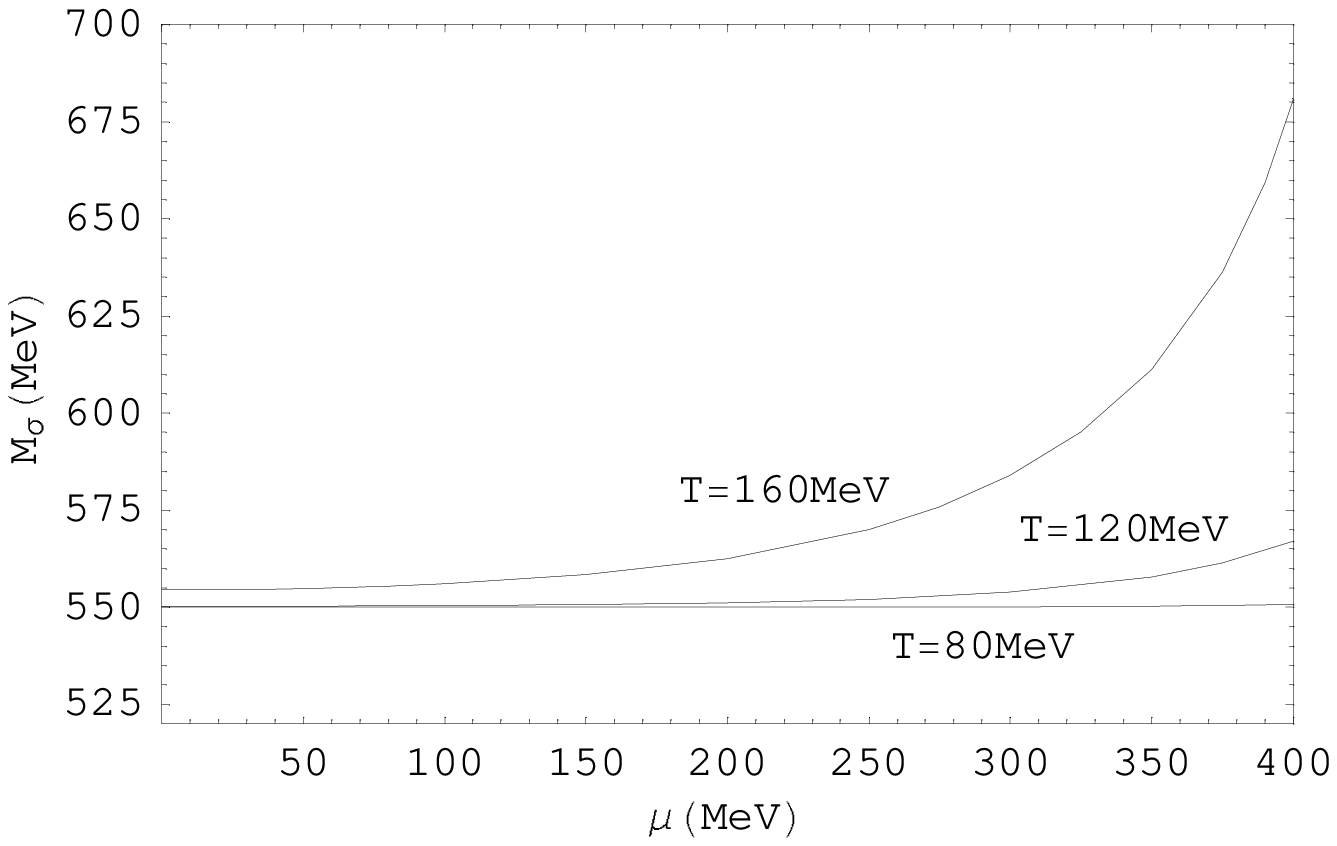}
\end{center}
\caption{The $\s$ effective masses as functions of chemical
potential at different given temperatures in the CJT formalism
.}\label{x7}
\end{figure}

\begin{figure}[tbh]
\begin{center}
\includegraphics[width=220pt,height=140pt]{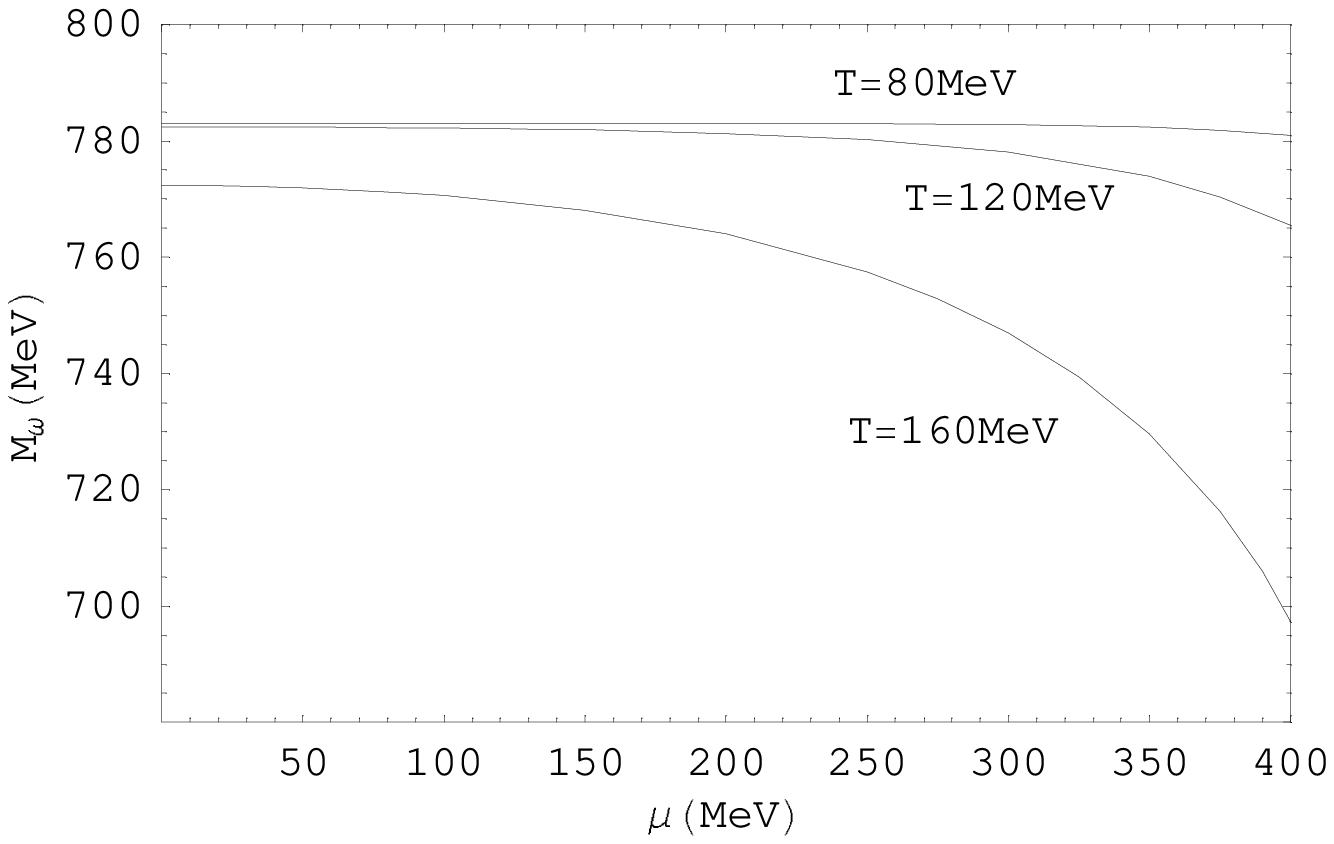}
\end{center}
\caption{The $\o$ effective masses as functions of chemical
potential at different given temperatures in the CJT formalism
.}\label{x8}
\end{figure}

We show in figures \ref{x7} and \ref{x8} the $\s$ and $\o$
effective masses as functions of chemical potential at different
give temperatures. The $\s$ effective masses increase with the
increase of the chemical potential while the $\o$ effective masses
decrease with the increase of the chemical potential at given
temperature. For both $\s$ and $\o$ mesons, the effective masses
change slightly with chemical potential at low given temperatures
but rapidly at high given temperatures.

\section{summary}
By the CJT formalism and Walecka model, we have made a beyond MFT
calculation in studying nuclear matter. We derive a coupled Dyson
equations for the full propagators of nucleons and mesons. After
neglecting the vacuum fluctuations and the momentum dependence in
the nucleon self-energies, the nucleon self-energies can be
evaluated by the Dyson equation of the nucleon full propagator,
while the meson effective masses are determined by minimizing the
thermodynamic potential to these effective masses instead of
directly evaluating the meson Dyson equations. This kind of
treatment simplifies the calculation meanwhile preserves the
thermodynamic consistency. The effective masses and thermodynamic
potential can be numerical evaluated at given chemical potential
and temperature. The numerical results show that the nucleon
effective mass decreases with the increase of temperature or
chemical potential, and it decreases more rapidly in the CJT case
than that of the MFT case. For the mesons, the effective mass of
$\o$ meson decreases with temperature or chemical potential
increasing, while the effective mass of $\s$ meson increases with
temperature or chemical potential increasing. Both nucleon and
meson effective masses change more quickly at high temperature or
chemical potential than that at low temperature or chemical
potential.

\begin{acknowledgments}
This work was supported in part by the National Natural Science
Foundation of China with No. 10547112 and No. 10675052.
\end{acknowledgments}

\end{document}